\documentclass{article}

\usepackage{arxiv}

\usepackage[utf8]{inputenc} 
\usepackage[T1]{fontenc}    
\usepackage{hyperref}       
\usepackage{url}            
\usepackage{booktabs}       
\usepackage{amsfonts}       
\usepackage{amsmath}
\usepackage{nicefrac}       
\usepackage{microtype}      
\usepackage{lipsum}
\usepackage{graphicx}
\graphicspath{ {./images/} }
\usepackage[numbers]{natbib}
\usepackage{subcaption}

\usepackage{xcolor}

\title{Improving Generative Inverse Design of Rectangular Patch Antennas with Test Time Optimization}

\author{
  Beck LaBash \\
  Northeastern University\\
  Boston, MA \\
  \texttt{labash.b@northeastern.edu} \\
  \And
  Shahriar Khushrushahi \\
  Notch Technologies\\
  Cambridge, MA \\
  \texttt{shahriar@notchtechnologies.com} \\
  \And
  Fabian Ruehle \\
  Northeastern University\\
  Boston, MA \\
  \texttt{f.ruehle@northeastern.edu} \\
}

\begin{document}
\maketitle

\begin{abstract}
We propose a two-stage deep learning framework for the inverse design of rectangular patch antennas. Our approach leverages generative modeling to learn a latent representation of antenna frequency response curves and conditions a subsequent generative model on these responses to produce feasible antenna geometries. 
We further demonstrate that leveraging search and optimization techniques at test-time improves the accuracy of the generated designs and enables consideration of auxiliary objectives such as manufacturability. Our approach generalizes naturally to different design criteria, and can be easily adapted to more complex geometric design spaces.

\end{abstract}

\section{Introduction}

In our increasingly wireless world, antennas serve as the fundamental link between radio frequency (RF) waves and electronic devices, and enable essential functions such as communication, navigation, and sensing across diverse systems, including GPS, WiFi, Bluetooth, and cellular networks. Among these, patch antennas—metallic patches printed onto dielectric substrates—are particularly attractive due to their low cost, low profile, and ease of fabrication. Rectangular patch antennas, in particular, are widely utilized in mobile devices and other space-constrained applications~\cite{balanis_antenna_theory_2016}.

For a patch antenna to be implemented for a specific use case, its behavior needs to be tuned to maximize efficiency. Specifically, the antenna should easily receive and transmit power at certain frequency ranges, while minimizing unwanted radiation or reception at other ranges. Each consisting of a center frequency and bandwidth, these frequency ranges characterize the frequency response of the antenna. Tuning the frequency response of the antenna has the effect of filtering out unwanted signals during reception, and maximizing power converted into radiation at the desired frequency during transmission.

However, designing patch antennas for optimal electromagnetic (EM) performance remains a time-consuming and iterative process. Engineers typically rely on combinations of analytic approximations and full-wave EM simulations —often based on numerical methods such as the Finite Element Method (FEM) or the Finite-Difference Time-Domain method (FDTD)— to match desired frequency responses \cite{jin_fem_2014, taflove_fdtd_2005}. While analytic models or lumped circuit approximations can guide preliminary geometric parameter selection, they are limited to simple geometries and often yield only coarse solutions \cite{balanis_antenna_theory_2016}. Additionally, adjusting one geometric parameter, such as the patch width, can simultaneously affect multiple aspects of the antenna’s frequency response, necessitating careful, often tedious manual tuning.

Deep learning design methods present a promising alternative by learning complex, nonlinear relationships directly from data and enabling a more automated and scalable inverse design process. Inverse design, in this context, involves starting from target EM behavior and determining the antenna geometry that produces it. Past breakthroughs in inverse design for other scientific domains, such as protein folding \cite{Jumper2021} and materials discovery \cite{merchant_materials_2023}, demonstrate the potential of deep learning to achieve results beyond conventional heuristics. Much like these domains, patch antenna design involves the navigation of a vast configuration space. Yet, we face notable challenges when applying deep learning in this context:

\begin{enumerate}
\item \textbf{Data Scarcity:} The absence of large, standardized datasets necessitates expensive simulation campaigns to train models effectively.

\item \textbf{Non-unique solutions:} The one-to-many nature of the inverse mapping —from a desired frequency response to multiple feasible geometries— further complicates direct regression approaches \cite{elmisilmani_ml_review_2020}.
\end{enumerate}

Addressing these obstacles requires specialized frameworks that balance multimodality and data efficiency.

In this work, we propose a novel two-stage generative framework for the inverse design of rectangular patch antennas that leverages variational autoencoders to model distributions over feasible electromagnetic responses and corresponding antenna geometries. The framework's adversarial training process enables controlled generation while its probabilistic nature addresses the one-to-many mapping challenge inherent to inverse design. Through our experiments, we show that simple search and optimization techniques employed at test time enhance design accuracy and practicability while reducing sensitivity to limited training data. Finally, while demonstrated specifically for rectangular patch antennas, we discuss how our approach naturally generalizes to arbitrary design criteria and more complex geometric design spaces.

\section{Related Work}

\paragraph{Patch Antenna Design}

Surrogate models serve as computationally inexpensive proxies for time-consuming EM simulations, enabling rapid evaluation of candidate designs. Early attempts employed simple linear regression or Support Vector Machines (SVMs) to approximate forward simulations of patch antennas \cite{chen_ml_observations_2022, chen_printed_patch_2022}. Although these models significantly accelerate the design process by replacing full-wave simulations, they still require iterative fine-tuning by human engineers, and their accuracy is limited by the complexity of the underlying EM phenomena. To fully automate the inverse design process, antenna geometries must be generated automatically from target frequency specifications. \citet{sharma_monopole_2020} constructed a surrogate-assisted approach to enumerate a fine grid of design parameters and selected geometries that met performance criteria. While effective, this approach suffers from poor scalability as the design space grows in dimensionality. Reinforcement learning (RL) has been proposed as a data-driven alternative to systematic search. \citet{wei_reinforcement_2023} formulated slot antenna design as a Markov Decision Process and trained a deep RL agent to iteratively refine geometric parameters until the target frequency response was achieved. The RL framework also integrated a surrogate model to limit expensive simulations, thereby overcoming both the data availability bottleneck and the one-to-many challenge of inverse design.

\paragraph{Controlled Generation}
Generative models often aim to produce outputs conditioned on specified target attributes. Variational Autoencoders (VAEs) \citep{kingma2013auto} are a popular class of generative models that learn a latent representation of data by encoding inputs into a latent space and then reconstructing them via a decoder. To enable controlled generation of samples with specified attributes, VAEs were extended to Conditional Variational Autoencoders (CVAEs) \citep{NIPS2015_8d55a249}, where an additional input to the decoder encodes the desired features of the output. However, the decoder can learn to ignore the conditional input if the latent code alone suffices to reconstruct the data. A solution to this problem is adversarial disentanglement, where an auxiliary predictor attempts to infer the condition from the latent code, while the encoder is penalized for revealing this information. This approach, employed in \citep{engel2019gansynth, makhzani2015adversarial}, encourages the model to encode all task-related information about the condition in the explicit conditional pathway, thus enabling controlled generation.

An alternative approach to controlled generation is to decouple the learning of the distribution from the conditioning altogether, training an unconditional Variational Autoencoder (VAE) on the data and then searching in its latent space to find a sample matching desired criteria \citep{Gómez-Bombarelli2018, nguyen2017plugplaygenerative}. This strategy avoids having to feed potentially out-of-distribution targets directly to the decoder; however, it places a greater computational burden on the search procedure and can become prohibitively expensive for large or high-dimensional search spaces.

In our work, we combine the benefits of both strategies through a two-stage generative pipeline. First, we train an unconditional VAE on physically valid frequency responses, allowing us to steer the latent code toward responses that remain in-distribution. Second, we train an adversarially disentangled CVAE to map those frequency responses to design parameters. This hybrid approach mitigates the pitfalls of purely conditional generation on out-of-distribution inputs, while also reducing the burden of exhaustive search in unconditional latent-space.

\paragraph{Test-Time Compute}
Test-time compute refers to the computational resources devoted to completing a task during inference. Sufficiently powerful test-time compute schemes can significantly boost performance, even when training data or model capacity is held constant. A seminal example of such an approach can be found in AlphaGo, which pioneered the concept of extensive test-time computation through Monte Carlo Tree Search (MCTS) combined with deep neural networks \cite{Silver2016-iz}. Recent success has been found in scaling test-time compute in large language model (LLM) systems to boost performance on different tasks. AlphaGeometry approached gold-medalist level performance on International Mathematical Olympiad geometry problems by extensively scaling beam search with a symbolic deduction engine for verification \cite{Trinh2024-hi}. Similarly, FunSearch made discoveries in open mathematics problems by combining systematic verification with evolutionary search through millions of programs \cite{Romera-Paredes2024-rs}. More recently, \cite{bonnet2024searchinglatentprogramspaces} demonstrated that test-time compute through latent program search can be leveraged to outperform strong priors (pretrained LLMs) on the Abstraction and Reasoning Corpus (ARC) \cite{chollet2019measureintelligence} benchmark. Building upon approaches that have worked well in photonics, we design our framework with test-time compute mechanisms to increase data efficiency and enable better generalization.

\section{Background}
\label{sec:background}
\begin{figure}[t]
    \centering
    \begin{subfigure}[b]{0.30\textwidth}
        \includegraphics[width=\textwidth]{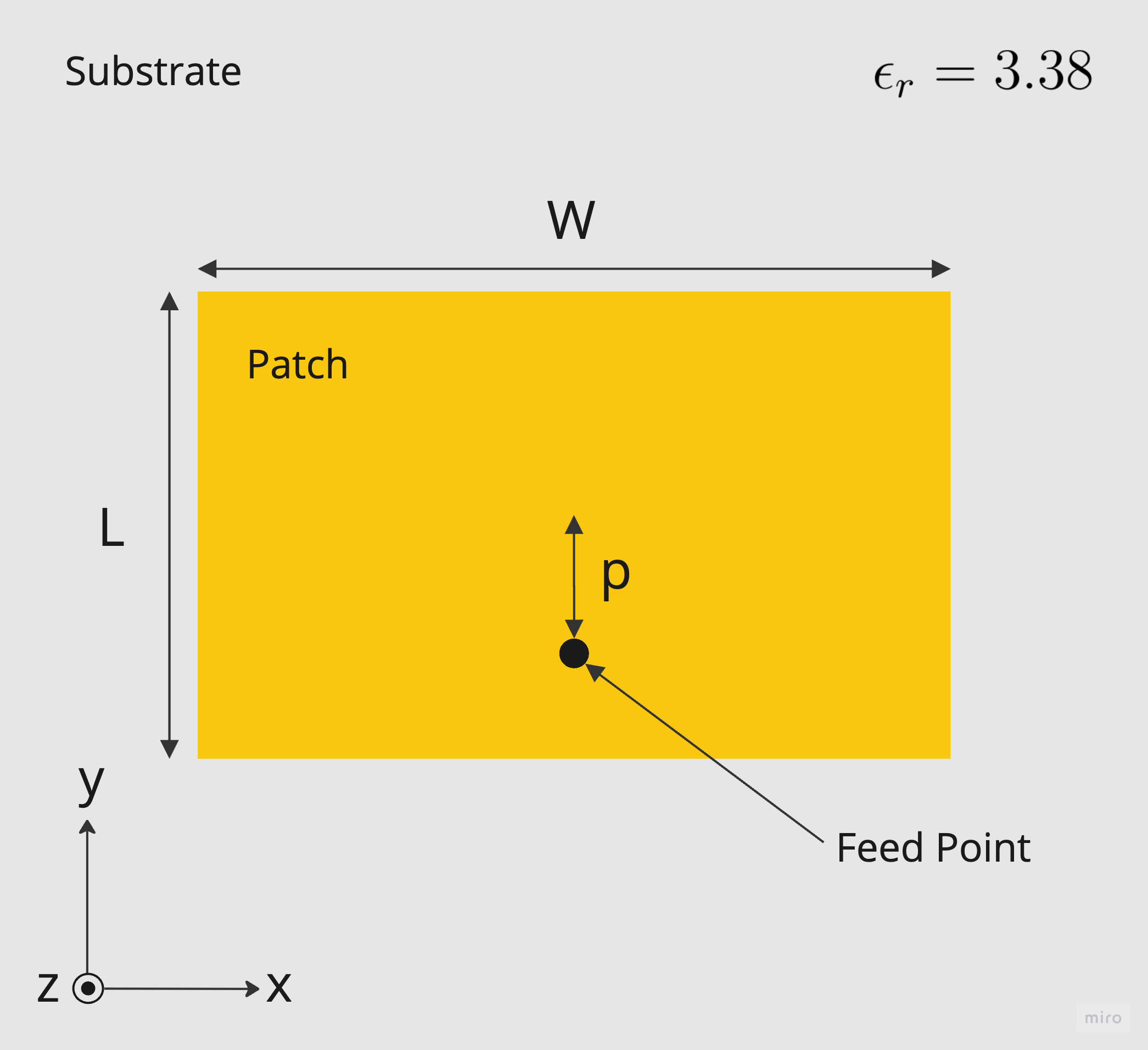}
        \caption{Top View}
    \end{subfigure}
    \hspace{0.8cm}
    \begin{subfigure}[b]{0.30\textwidth}
        \includegraphics[width=\textwidth]{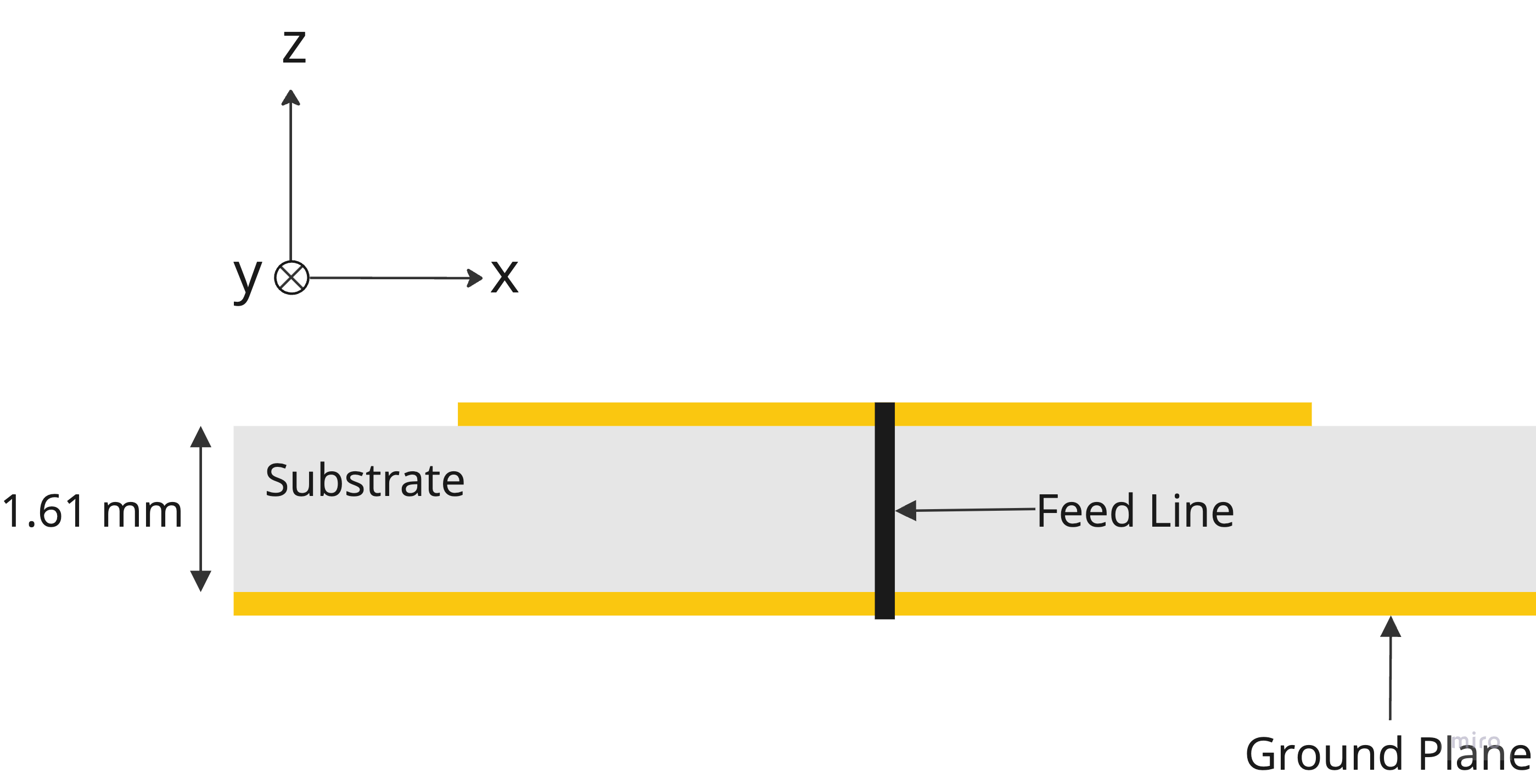}
        \caption{Side View}
    \end{subfigure}
    \caption{Configuration of a Rectangular Patch Antenna fed via coaxial line through the ground plane.}
    \label{fig:patch_spec}
\end{figure}

In this work, we focus on a coaxial-fed rectangular patch, where a feed pin passes through a ground plane to a metallic patch on a dielectric substrate (cf.\ Figure~\ref{fig:patch_spec}). The design configuration of a single coaxial-fed rectangular patch antenna is parametrized by $(L, W, p)$, where $L$ is the length of the patch in mm, $W$ is the width of the patch in mm, and $p$ is the distance of the feed point from the center of the patch along the length axis. 

The patch can be treated as a resonant segment of a transmission line, with \(L\) primarily controlling the resonant frequency and \(W\) influencing impedance and bandwidth. The feed position \(p\) is tuned to achieve an impedance match (commonly \(50\,\Omega\)) with the feed at a desired resonant frequency. Our primary goal is to generate antenna geometries \((L, W, p)\) that satisfy specific frequency-domain performance requirements, encoded in the reflection coefficient \(S_{11}(f)\). 

\section{Dataset}

We build a dataset of 1292 design configurations and corresponding $S_{11}$ frequency response curves to train our inverse design framework. 

To generate a set of $(L, W, p)$ to be simulated, we begin by computing a grid of configurations with the bounds $L = [7.5,\; 52.5]$, $\frac{W}{L}\text{ ratio} = [0.8,\; 2]$, and $p = [-6,\; 0)$ while enforcing $p =  (-\frac{L}{2},\; 0)$ and sampling at higher density at small $L$ and $p$ close to the edge of the patch. We then augment this initial set of designs by using an algorithm designed to sample additional triplets $(L, W, p)$ inside the convex hull of the existing dataset while enforcing uniformity. Ultimately, we end up with a set of 1292 design configurations.

To simulate the designs, we use openEMS \cite{openEMS}, an open source electromagnetic field solver based on the Finite-Difference Time Domain (FDTD) method. We fix substrate parameters $\epsilon_r = 3.68$ and $\text{thickness} = 1.61 \text{mm}$ to align with those provided by OSH Park's 4 layer prototype service. To calculate $S_{11}$ frequency response curves, we excite the antenna with a Gaussian pulse centered at $f_0 = 5.5\,\text{GHz}$ and a bandwidth defined by a cutoff frequency $f_c = 4.5\,\text{GHz}$ to cover the frequency range of interest, $f\in[1\text{GHz},\; 10\text{GHz}]$. 

From the port data extracted through simulation, we obtain the complex amplitudes of the incident and reflected fields at $N = 1000$ regularly spaced frequencies in this range and compute the reflection coefficient ($S_{11}$) as the ratio of the reflected wave ($u_{\text{ref}}$) to the incident wave ($u_{\text{inc}}$), converted to decibels,

\begin{align}
|S_{11}(f_i)|_\text{dB} = 20\log_{10}\left(\left|\frac{u_{\text{ref}}(f_i)}{u_{\text{inc}}(f_i)}\right|\right),\quad i=1,\dots,N
\label{eq:dBConversion}
\end{align}

which is stored for each antenna design configuration in the dataset
\begin{align*}
\mathcal{D}=\left\{\bigl((L,W,p)^{(j)},y^{(j)}\bigr)\right\}_{j=1}^{1292}, 
\quad\text{with } y^{(j)}=\left[|S_{11}(f_1)|_\text{dB},\dots,|S_{11}(f_N)|_\text{dB}\right]^\top \in\mathbb{R}^{N}.
\end{align*}

\section{Methodology}
\label{sec:methodology}

\begin{figure}[t]
    \centering
    \includegraphics[width=\textwidth]{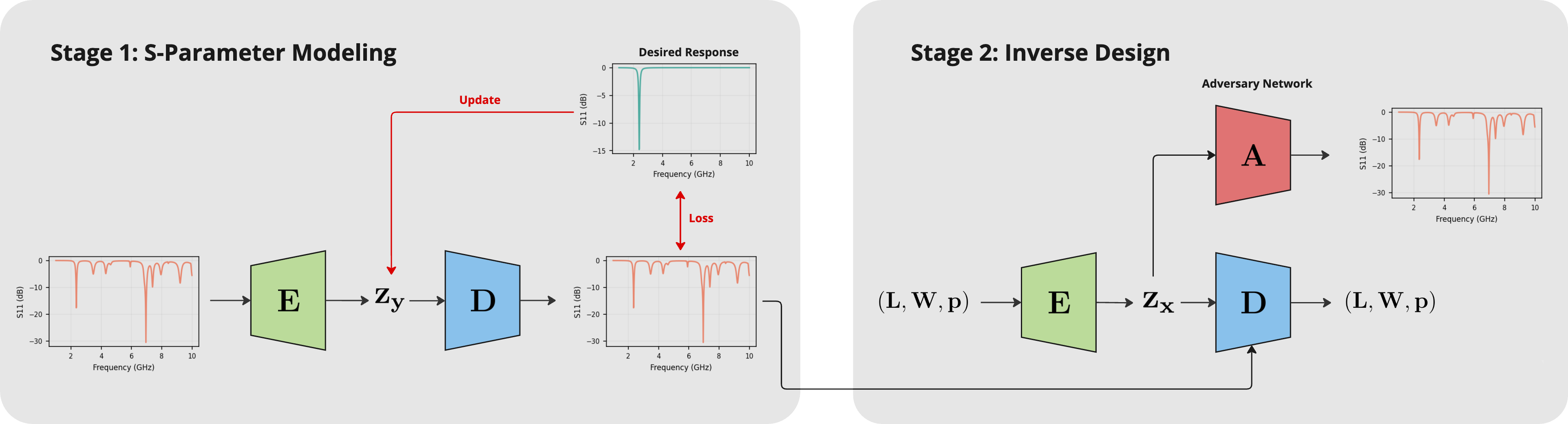}
    \caption{Overview of our two-stage generative inverse design framework. Stage 1 learns a latent representation of $S_{11}$ frequency response curves and finds in-distribution curves matching target responses. Stage 2 uses a conditional VAE to generate antenna geometries that produce the desired EM response. Red arrows represent test time optimization.}
    \label{fig:framework}
\end{figure}

\subsection{Problem Setup and Overview}
We consider antennas defined by \( x = (L, W, p) \), where \( L \) and \( W \) are the patch dimensions and \( p \) is the feed position. Each design \( x \) yields a frequency-dependent reflection coefficient \( S_{11}(f) \), sampled at \( N=1000 \) frequencies. We represent this response as a 1000-dimensional vector \( y \in \mathbb{R}^N \), where we convert the response curve to dB, see Equation~\eqref{eq:dBConversion}.

We define our design criteria as a set of $K$ desired resonant frequencies $f_\text{res} = \{f_1, \ldots, f_K\}$ with their corresponding bandwidths $BW = \{BW_1,\ldots, BW_K\}$ and depths $d = \{d_1, \ldots, d_K\}$. A lower \(|S_{11}|_\text{dB}\) near a desired frequency band indicates efficient power transfer and radiation, whereas higher \(|S_{11}|_\text{dB}\) elsewhere can mitigate interference. This means we are searching for antenna that have an idealized target response curve modeled as a product of Lorentzian notches,
\begin{align}
\label{eq:LorentzianNotch}
    S_{11}^*(f~|~ f_\text{res}, BW, d) = \prod_{k=1}^{K} \left[ 1 - (1-10^{\frac{d_k}{20}})\left(1 - \left(1 - \frac{(BW_k/2)^2}{(f - f_k)^2 + (BW_k/2)^2}\right)\right) \right]\,,
\end{align}
where \(f_k\) denotes the resonant frequency where minimal reflection is desired, \(BW_k\) specifies the width of the frequency band around \(f_k\) with efficient power transfer, and \(d_k\) sets the target depth (in dB) of the reflection at \(f_k\). For example, choosing parameters \((f_k=2.4\,\text{GHz}, BW_k=0.2\,\text{GHz}, d_k=-15\,\text{dB})\) describes an antenna with efficient radiation around 2.4 GHz and reflection quickly increasing outside this range.

From this response curve, we convert to idealized target dB values
\begin{align}
\label{eq:targetY}
y^*(f_i~|~ f_\text{res}, BW, d) = 20 \log_{10}|S_{11}^*(f_i~|~ f_\text{res}, BW, d)|, \quad i=1,\dots N\,.
\end{align}

Of course, antennas with these exact frequency response curves do not exist in practice (at the very least, there will be higher harmonics). Thus, we are searching for antenna designs whose frequency response curve has a depth $y(f_k)$ of at least $d_k$ in a frequency band around $f_k$, and whose response remains shallower than a threshold for all other (non-resonant) frequencies. This means we will not consider $y$-values deeper than this threshold,  in frequency ranges that are irrelevant to our target response in our search.

Our inverse design goal is as follows: given an idealized target response $y^*(f_\text{res}, BW, d)$, find antenna design parameters $\tilde{x}$ such that $\varphi_\text{EMS}(\tilde{x}) \simeq y^*$, where $\varphi_\text{EMS}$ represents the forward EM simulation~\cite{openEMS}.
Since $\varphi_\text{EMS}$ is expensive to evaluate and the preimage $y^* \mapsto \tilde{x}$ is not unique, we propose a two-stage generative pipeline (see Fig.~\ref{fig:framework}):

\begin{enumerate}
    \item \textbf{Stage 1: Latent Representation of $\boldsymbol{S_{11}}$-Responses.} 
    We train a VAE that encodes $y \in \mathbb{R}^{N}$ into a latent vector $z_y \in \mathbb{R}^{64}$, capturing the key variations of physically realizable frequency responses. At test time, we search the latent space of this VAE to find an approximation $\tilde{y}$ of $y^*(f_\text{res}, BW, d)$.
    
    \item \textbf{Stage 2: Conditional Design Generation.} 
    We train a conditional VAE (CVAE) whose encoder maps $x\in\mathbb{R}^3$ to $z_x\in\mathbb{R}^{16}$, and a decoder that maps $z_x$ conditioned on the desired frequency response $\tilde{y}$ to a design $\tilde{x}$. To ensure $\tilde{x}$ depends both on $z_x$ and $\tilde{y}$ (rather than just $z_x$), we use an adversarial predictor that discourages $z_x$ from leaking information about $\tilde{y}$. At test time, we use this CVAE to decode $\tilde{x}$ from $\tilde{y}$ and $z_x$.
\end{enumerate}

We further refine the accuracy of the generated design through (1) best-of-$N$ sampling of $z_y$, $z_x$, and (2) gradient-based optimization of each $z_x$ based on secondary constraints, e.g., physical feasibility or specific manufacturing limits, without losing the desired EM response.

\subsection{Latent Representation of \texorpdfstring{$\boldsymbol{S_{11}}$}{S11}-Responses}
To enable search through the distribution of feasible \( S_{11} \) frequency response curves, we train a Variational Autoencoder \citep{kingma2013auto} (VAE) on our dataset. The VAE defines
\[
z_y \sim q_\phi(z_y|y), \quad y \sim p_\theta(y|z_y)\,,
\]
where \( q_\phi(z_y|y) \) is the approximate posterior (encoded by convolutional layers) and \( p_\theta(y|z_y) \) is the likelihood (decoded by transposed convolutional layers). We assume a Gaussian prior \( p(z_y)=\mathcal{N}(0,I) \).

We employ a \(\beta\)-VAE style objective \citep{higgins2017beta}, with \(\beta<1\) chosen to strike the best generative quality in our experiments,
\[
\mathcal{L}_{\text{VAE}} = \mathbb{E}_{q_\phi(z_y|y)}[\log p_\theta(y|z_y)] - \beta\,D_{\text{KL}}(q_\phi(z_y|y)\|p(z_y))\,.
\]

This encourages a structured latent space capturing essential variations in physically realizable \( S_{11} \)-responses.

\subsection{Latent Search for Characteristic Target Responses} \label{sec:s11search}
Using the idealized target response $y_*$ computed from~\eqref{eq:LorentzianNotch} via~\eqref{eq:dBConversion} in the CVAE can lead to undesirable results as it is potentially out of the distribution of realizable $y$. Instead, we search the VAE latent space to find latent variables $z_{y}$ that decode to $\tilde{y}\approx y$ which has the same depths at frequency bands as our idealized target $y^*$, but can differ from $y^*$ in regions we do not care about in our design goals (as mentioned above, antennas producing the idealized frequency response curve do not exist). To find good latent space vectors, we perform a gradient-based search over the latent space $z_y$. We choose the starting point $z_y^{(0)}$ using one of two strategies (random or nearest neighbor), which we explore further in Section~\ref{sec:TestTimeScaling}, and define the objective
\begin{align*}
z_y^{(t+1)} = z_y^{(t)} - \alpha \nabla_{z_y}\bigl(\| p_\theta(y|z_y) - y^{*}\|_{\text{masked}}^2 + \lambda_{\text{reg}}\|z_y\|^2 \bigr)\,,  
\end{align*}
where \(\alpha\) is the learning rate. In the loss, we mask $y$ in frequency regions that are irrelevant for our desired response curves, ensuring we only penalize differences in regions of interest. We also add a regularization term $\lambda_{\text{reg}}\|z_y\|^2$ to keep the solution near the prior. This iterative optimization yields a $z_y^{*}$ whose decoded response $\tilde{y}$ closely approximates  $y^{*}$ and remains on the manifold of physically realizable curves.

\subsection{Controlled Design Generation}
To generate antenna designs \( x \) with a frequency response \( y \), we train a conditional VAE (CVAE):
\[
z_x \sim q_\varphi(z_x|x), \quad x \sim p_\psi(x|z_x,y)\,,
\]
where \( q_\varphi \) and \( p_\psi \) are feed-forward networks mapping between design parameters and latent variables. 

However, during training, \(z_x\) may easily learn to encode all information about $x$ needed for reconstruction. As noted by \cite{engel2019gansynth}, this can lead to a pathological scenario where the decoder simply ignores the conditional input $y$, since the necessary information is already present in the latent code $z_x$. In such cases, attempts to control generation by varying $y$ would have no effect.

To prevent this, we follow \cite{engel2019gansynth} and introduce an adversarial predictor \( D_\omega \) that attempts to predict \( y \) directly from \( z_x \). We let $\eta$ be a hyperparameter controlling the weight of the adversarial term. The predictor tries to minimize
\[
\mathcal{L}_{\text{pred}} = \mathbb{E}_{q_\varphi(z_x|x)}[\|y - D_\omega(z_x)\|^2]\,.
\]

The encoder tries to maximize this error (i.e., make it hard to predict \( y \) from \( z_x \)). Incorporating this into the CVAE training, we obtain a combined objective:
\[
\mathcal{L}_{\text{CVAE,combined}} = \mathbb{E}_{q_\varphi(z_x|x,y)}[\log p_\psi(x|z_x,y)] - \beta_x D_{\text{KL}}(q_\varphi(z_x|x,y)\|p(z_x)) - \eta \mathcal{L}_{\text{pred}}\,,
\]
where \(\mathcal{L}_{\text{pred}}\) is the adversarial term from the encoder’s perspective (the encoder seeks to increase \(\mathcal{L}_{\text{pred}}\), the predictor seeks to decrease it). By this minimax interplay, the encoder removes direct correlation between \( z_x \) and \( y \).

Since \( x \) maps to a unique \( y \) via the EM solver, forcing the encoder to remove \( y \)-information from \( z_x \) ensures the decoder must rely on the explicit conditional input \( y \) to reconstruct the design. This yields a controllable model: changes in \( y \) at test time directly influence \( x \), and \( z_x \) can be adjusted to handle auxiliary objectives without altering the frequency response of \( x \).

\subsection{Test Time Optimization}

While our two-stage framework generates feasible designs directly from target specifications, additional refinement at test time can further improve accuracy and practicability. We consider two approaches: (1) generating multiple candidate designs and (2) optimizing single candidates to satisfy auxiliary constraints.

\paragraph{Generating Multiple Candidates.} By sampling multiple response curves $\tilde{y}$ during latent search and multiple designs from the conditional VAE for each candidate, we obtain a pool of potential solutions. 
We employ two scoring methods to rank generated design candidates at test time:

\emph{Oracle Scorer:} Runs a full EM simulation and computes an MSE against the target $y^{*}$. Although accurate, this approach is computationally expensive.

\emph{Surrogate Scorer:} Uses a neural network trained with a $\beta$-NLL loss \citep{seitzer2022pitfallsheteroscedasticuncertaintyestimation} to approximate the forward simulation. At test time, we incorporate the surrogate's uncertainty estimates into our score, adjusting the influence of different frequency regions based on the model's predictive confidence.

Both scoring methods consider only frequency regions relevant to our target response. Regions outside the specified resonant frequency bands are masked, meaning their errors do not contribute to the score.

\paragraph{Optimizing a Single Candidate.} For a single candidate curve, the latent code of the conditional design decoder can be further optimized to meet auxiliary geometric criteria. For the rectangular patch antenna, we define a penalty
\[
\mathcal{L}_\text{penalty}(L, W, p) = \text{ReLU}(-L)^2 + \text{ReLU}(-W)^2 + \text{ReLU}\left(-\frac{L}{2} - p\right)^2 + \text{ReLU}(p)^2\,,
\]
encouraging positive dimensions and a feed position between the center and edge of the patch. We then use this penalty to conduct a gradient-based search over the latent variable $z_x$ in the same way as described in Section~\ref{sec:s11search} for $z_y$.

\section{Experiments}

\subsection{Model Architecture and Training Details}

\paragraph{Stage 1 (VAE).}
We use a convolutional Variational Autoencoder (VAE) to map 1000-dimensional frequency response curves into a 64-dimensional latent space. The encoder has five convolutional layers with ReLU activations, while the decoder has four transposed convolutional layers with GELU activations, dropout, and batch normalization. The VAE is trained for 250 epochs (batch size=32) using Adam (lr=$10^{-3}$), with a KLD weight of 0.016 annealed over the first 100 epochs.

\paragraph{Stage 2 (CVAE).}
The Conditional Variational Autoencoder (CVAE) maps antenna parameters to a 16-dimensional latent space using fully-connected layers: two linear layers with LeakyReLU activations for the encoder, and five linear layers with GELU activations for the decoder. A convolutional head (reusing Stage 1's encoder) extracts embeddings from frequency response curves to condition the generation. An adversarial predictor (reusing Stage 1's decoder with an additional fully connected layer to map the inputs) enforces conditional dependence, using a disentanglement weight of 0.1. The CVAE is trained for 300 epochs (batch size=32) using Adam (lr=$10^{-3}$), with a KLD weight of 0.016 annealed over the first 50 epochs.

\paragraph{Surrogate Scorer.}
The surrogate scorer reuses the decoder architecture from Stage 1, modified to have two output channels (mean and variance), and employs heteroscedastic Gaussian likelihood ($\beta$-NLL with $\beta=0.5$) to quantify predictive uncertainty. It is trained for 500 epochs (batch size=64) using the Adam optimizer (lr=$5\times10^{-3}$).

\paragraph{Computational Resources and Training Time.}
All models are implemented in PyTorch using the Metal Performance Shaders (MPS) backend and trained on an Apple M2 chip. Training times per model are: Stage 1 VAE (15 min), Stage 2 CVAE (5 min), and Surrogate Scorer (15 min). Experiments use model checkpoints selected based on optimal validation performance.

\subsection{Test Time Compute Scaling}
\label{sec:TestTimeScaling}

\begin{figure}[t]
    \centering
    \begin{subfigure}[b]{0.45\textwidth}
        \includegraphics[width=\textwidth]{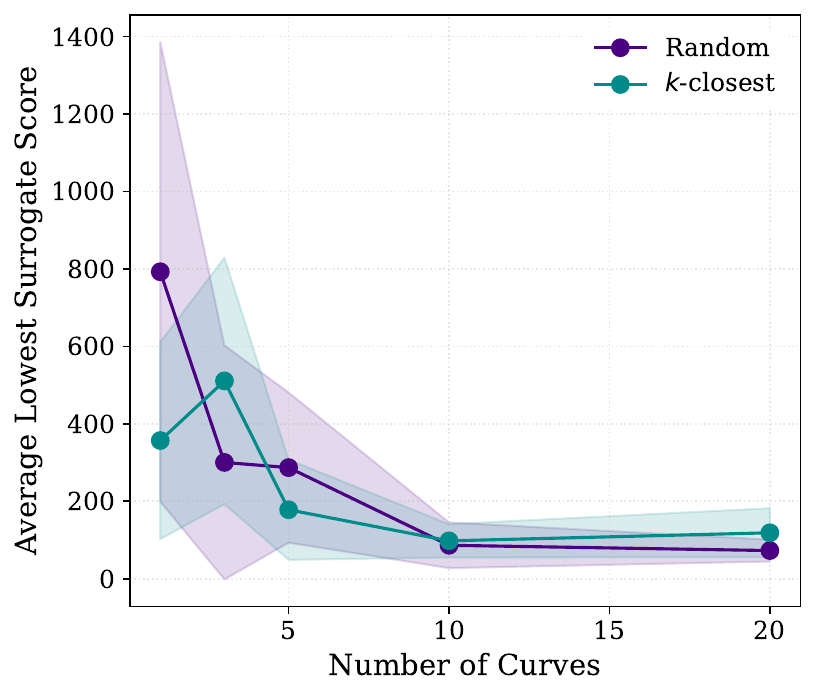}
        \caption{Performance vs. Number of Curves}
    \end{subfigure}
    \hspace{0.05\textwidth}
    \begin{subfigure}[b]{0.45\textwidth}
        \includegraphics[width=\textwidth]{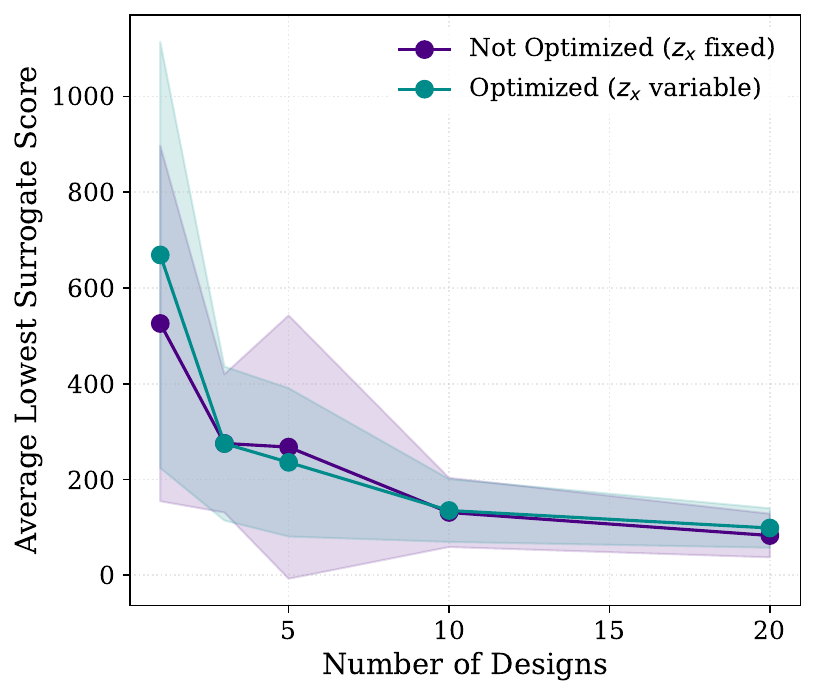}
        \caption{Performance vs. Number of Designs Sampled per Curve}
    \end{subfigure}
    \caption{Scaling performance as the number of curves (left) and the number of designs per curve (right) is increased. The shaded regions indicate variability across runs.}
    \label{fig:scaling_graphs}
\end{figure}

We conduct two investigations to determine how increasing test-time compute affects framework performance. In both investigations, we consider three distinct target response functions, and at each search configuration, record the lowest \emph{Surrogate Scorer} score from the pool of generated antenna designs, averaged across the three targets. Results can be seen in Figure~\ref{fig:scaling_graphs}.

\paragraph{Multiple Candidate Curves.} To determine how sampling multiple candidate frequency response curves $\tilde{y}$ affects design accuracy, we consider a setup where one antenna design $\tilde{x}$ is sampled per candidate curve, and vary the number of candidate curves. To obtain the different curves $\tilde{y}$, we explore two initialization strategies for $z_y^{(0)}$: 
\begin{enumerate}
    \item \textbf{Random initialization} samples each $z_y$ from a Gaussian prior 
    \item \textbf{$\boldsymbol{k}$-closest initialization} initializes $z_y$ as the latent codes of the $k$ curves in the training dataset that are closest to the ideal curve ($k$ is the number of candidate curves).
\end{enumerate}
We find that as the pool of antenna designs $\tilde{x}$ grows through the amount of candidate frequency response curves, both initialization strategies yield better and more consistent predictions.
Additionally, we find that early on, $k$-closest initialization may offer more stable or slightly better performance than random initialization, but as the number of curves increases, the difference in their average performance diminishes. This makes sense, since $k$-closest initialization potentially offers a more principled starting point for optimization in the one-shot case. 

\paragraph{Multiple Candidate Designs per Curve.} To determine how the number of generated designs per candidate curve affects design accuracy, we consider a setup where only one candidate response curve  $\tilde{y}$ is sampled, and vary the number of designs sampled from this curve. Additionally, we explore the case where $z_x$ is optimized according to the auxiliary geometric criteria and compare it to the unoptimized performance.

Again, we find that the quality of the antennas improve as we sample multiple  antenna designs $\tilde{x}$. Additionally, we find that the quality of the optimized and unoptimized design agrees well across search configurations, indicating that we have been successful in decorrelating $z_x$ from the physical response of $x$.

\subsection{Comparison of Target vs. Simulated Responses}

\begin{figure}[t]
    \centering
    \includegraphics[width=0.6\textwidth]{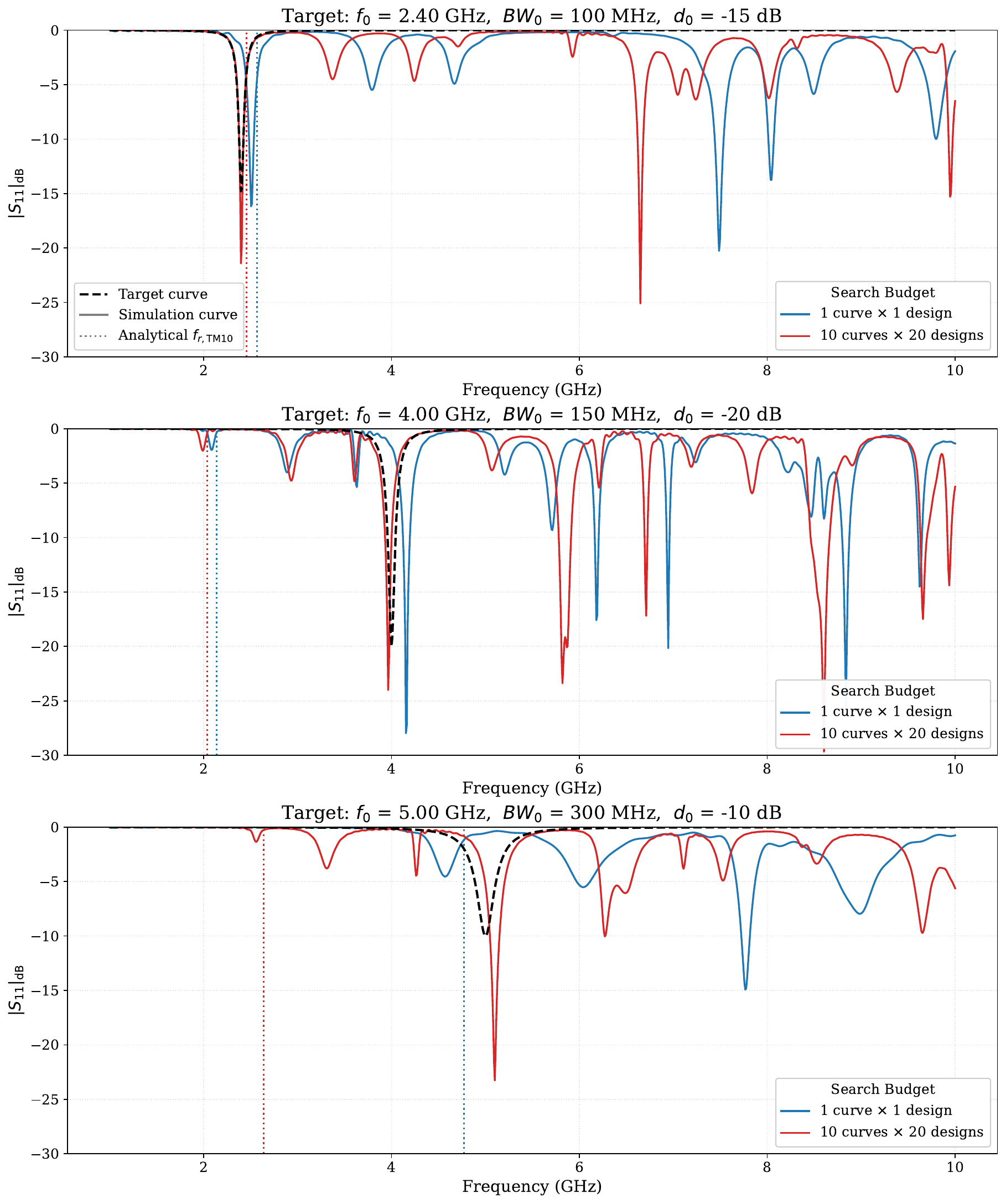}
    \caption{%
        Comparison of the idealised target $S_{11}$ curve $y^{*}$ (black dashed),
        the dominant-mode analytic resonance $f_{r,\mathrm{TM10}}$ (dotted vertical),
        and the simulated $S_{11}$ of designs $\tilde{x}$ generated with two
        test-time compute budgets.
        Blue = 1 curve $\times$ 1 design, red = 10 curves $\times$ 20 designs.  
        \textit{Note:} Generated geometries may exploit higher-order or coupled
        modes, so exact agreement with the analytic reference is not expected.%
    }
    \label{fig:s11_response_comparison}
\end{figure}

\begin{table}[t]           
\centering
\caption{Patch-antenna geometries corresponding to the frequency
responses in Figure~\ref{fig:s11_response_comparison}.}
\label{tab:geometry}
\begin{tabular}{@{}lcccc@{}}
\toprule
Target $f_r$ & Budget (\# curves $\times$ \# designs) & $L$ [mm] & $W$ [mm] & $p$ [mm]\\
\midrule
2.40 GHz & 1 $\times$ 1  & 30.0 & 40.5 & $-7.4$  \\
                   & 10 $\times$ 20 & 31.3 & 45.8 & $-7.2$ \\[2pt]
4.00 GHz & 1 $\times$ 1  & 36.0 & 53.6 & $-2.6$ \\
         & 10 $\times$ 20 & 37.8 & 52.7 & $-2.7$ \\[2pt]
5.00 GHz & 1 $\times$ 1  & 15.7 & 25.5 & $-2.1$ \\
         & 10 $\times$ 20 & 29.0 & 46.9 & $-1.7$ \\
\bottomrule
\end{tabular}
\end{table}

To illustrate the accuracy of the framework, Figure~\ref{fig:s11_response_comparison} compares simulated $S_{11}$ responses against the idealized target curves for antenna designs generated under two search conditions: (1) a single candidate curve is generated, from which a single design is sampled and (2) 10 candidate curves are generated, from which 20 design curves are sampled (pool of 200), and the design with lowest \emph{Surrogate Scorer} score is chosen. In both cases, we use random latent initialization and do not optimize individual designs according to the manufacturability constraint. 

The figure shows examples at $f_0=2.4$ GHz, $f_0=4.0$ GHz, and $f_0=5.0$ GHz with varying bandwidths and depths. In each case, it seems that devoting more compute to search over the design space at test time yields a more accurate result. For every generated design we plot the dominant-mode analytic resonance $f_{r,\mathrm{TM10}}$ computed with the Hammerstad–Jensen correction \cite{Hammerstad1975EquationsFM}.  While agreement with the target resonance is seen in the $f_0=2.40$ GHz example, the search occasionally favors geometries whose performance hinges on higher-order or coupled modes rather than the textbook TM$_{10}$ resonance.

In general, the framework is able to produce a design that closely matches the desired response. However, the limitations of the framework can be observed in the $f_0=5.0$ GHz curve where the design fails to meet the depth ($d = -10$) dB requirement even after an extensive search.

\section{Conclusion}
In this work, we presented a two-stage generative framework for the inverse design of rectangular patch antennas that effectively addresses key challenges in computational antenna design. Our approach combines the strengths of generative modeling with targeted test-time optimization to yield physically realizable antenna designs that meet desired frequency response characteristics.

Our experimental results demonstrate that test-time computation can dramatically improve design quality with minimal additional training data or model complexity. As shown in Section 6, both best-of-$N$ sampling and gradient-based optimization at test time yield progressively better designs that more closely match target frequency responses. This finding aligns with recent successes in other domains where scalable test-time computation has unlocked performance improvements without requiring larger models or datasets.

Our approach generalizes naturally to more complex electromagnetic design tasks. While we focused on rectangular patch antennas with three design parameters, the same framework could be extended to patch antennas with arbitrary geometries or to multi-element antenna arrays with more complex frequency-domain behavior. Furthermore, our test-time optimization framework offers a flexible mechanism for incorporating auxiliary design constraints, such as fabrication limitations or size restrictions, without compromising the primary electromagnetic performance objectives.

\section*{Code and Data Availability}
We have made our dataset, simulation scripts, trained models, and source code to reproduce all experiments and figures publicly available at \url{https://github.com/becklabs/patch-antenna-tto}.

\bibliographystyle{plainnat}  
\bibliography{references}

\end{document}